\documentclass[prd,twocolumn,showpacs,amsmath,amssymb,nofootinbib]{revtex4-1}
\usepackage{comment}
\usepackage{makecell}
\usepackage{graphicx}
\usepackage{bm}
\usepackage{color}
\usepackage{overpic}
\usepackage{hyperref}
\usepackage[mathlines]{lineno}
\usepackage{etoolbox,xspace}

\newcommand{\CP}{{\it CP}\xspace}

\newcommand{\alphaXi}{\alpha_{\Xi}}
\newcommand{\alphaz}{\alpha_{\Lambda 0}}
\newcommand{\baralphaz}{\bar{\alpha}_{\Lambda 0}}
\newcommand{\alpham}{\alpha_{\Lambda -}}
\newcommand{\alphap}{\alpha_{\Lambda +}}

\begin{document}

\title{\boldmath Investigation of the $\Delta I = 1/2$ rule and test of CP symmetry through 
the measurement of decay asymmetry parameters in $\Xi^-$ decays
}

\author{ 
M.~Ablikim$^{1}$, M.~N.~Achasov$^{5,b}$, P.~Adlarson$^{75}$, X.~C.~Ai$^{81}$, R.~Aliberti$^{36}$, A.~Amoroso$^{74A,74C}$, M.~R.~An$^{40}$, Q.~An$^{71,58}$, Y.~Bai$^{57}$, O.~Bakina$^{37}$
% ...
\\
\vspace{0.2cm}
(BESIII Collaboration)\\
\vspace{0.2cm}{\it
$^{1}$ Institute of High Energy Physics, Beijing 100049, People's Republic of China\\
% ...
$^{5}$ Budker Institute of Nuclear Physics SB RAS (BINP), Novosibirsk 630090, Russia\\
% ...
$^{36}$ Johannes Gutenberg University of Mainz, Johann-Joachim-Becher-Weg 45, D-55099 Mainz, Germany\\
$^{37}$ Joint Institute for Nuclear Research, 141980 Dubna, Moscow region, Russia\\
% ...
$^{40}$ Liaoning Normal University, Dalian 116029, People's Republic of China\\
% ...
$^{57}$ Southeast University, Nanjing 211100, People's Republic of China\\
$^{58}$ State Key Laboratory of Particle Detection and Electronics, Beijing 100049, Hefei 230026, People's Republic of China\\
% ...
$^{71}$ University of Science and Technology of China, Hefei 230026, People's Republic of China\\
% ...
$^{74}$ University of Turin and INFN, (A)University of Turin, I-10125, Turin, Italy; (B)University of Eastern Piedmont, I-15121, Alessandria, Italy; (C)INFN, I-10125, Turin, Italy\\
$^{75}$ Uppsala University, Box 516, SE-75120 Uppsala, Sweden\\
% ...
$^{81}$ Zhengzhou University, Zhengzhou 450001, People's Republic of China\\
\vspace{0.2cm}
% ...
$^{b}$ Also at the Novosibirsk State University, Novosibirsk, 630090, Russia\\
% ...
}}
\date{\today}
\begin{abstract}

Using $(10087\pm44)\times 10^{6}$ $J/\psi$ events collected with the
BESIII detector, numerous $\Xi^-$ and $\Lambda$ decay asymmetry
parameters are simultaneously determined from the process $J/\psi \to
\Xi^- \bar{\Xi}^+ \to \Lambda(p\pi^-) \pi^- \bar{\Lambda}(\bar{n}
\pi^0) \pi^+$ and its charge-conjugate channel. The precisions of
$\alphaz$ for $\Lambda \to n\pi^0$ and $\baralphaz$ for $\bar{\Lambda}
\to \bar{n}\pi^0$ compared to world averages are improved by factors
of 4 and 1.7, respectively. The ratio of decay asymmetry parameters of
$\Lambda \to n\pi^0$ to that of $\Lambda \to p\pi^-$, $\langle \alphaz \rangle/ \langle \alpham \rangle $, is
determined to be $ 0.873 \pm 0.012^{+0.011}_{-0.010}$, where the first
and the second uncertainties are statistical and systematic,
respectively. 
The ratio is smaller than unity more than $5\sigma$,
which signifies the existence of the $\Delta I = 3/2$ transition
in $\Lambda$  for the first time.
Besides, we test for CP symmetry in $\Xi^- \to \Lambda \pi^-$ and 
in $\Lambda \to n \pi^{0}$ with the best precision to date.

\end{abstract}

\maketitle

The non-invariance of fundamental interactions under the combination
of charge-conjugation (C) and parity (P) transformations is a
necessary condition for baryogenesis~\cite{Sakharov:1967dj}, a process
that dynamically generates the matter-antimatter asymmetry in the
universe. Although the Standard Model (SM) accommodates CP violation
with the Kobayashi-Maskawa
phase~\cite{Cabibbo:1963yz,Kobayashi:1973fv}, it can only explain a
matter-antimatter asymmetry that is at least ten orders of magnitude
smaller than the observed value~\cite{Barr:1979ye}.  Additional
sources of CP violation beyond the SM are expected to exist, and the
weak hadronic transitions of hyperons are another place to search
for such sources of CP
violation~\cite{Salone:2022lpt,Donoghue:1986hh}.

When two or more transition amplitudes interfere with each other,
relative weak- and strong-phase contributions exist between them.
For $K\to\pi\pi$~\cite{KTeV:1999kad, NA48:1999szy}, the CP violating weak
phase comes from the interference between $S$-wave isospin
$I=0$~($A_{0}$) and isospin $I=2$~($A_{2}$) amplitudes, which
correspond to $\Delta I=1/2$ and $\Delta I=3/2$ transitions,
respectively~\cite{Cirigliano:2011ny}.  The unforeseen large discrepancy
between the real parts of the two isospin amplitudes, ${\rm
Re}(A_{0})/{\rm Re}(A_{2})=22.45 \pm 0.06$, known as the $\Delta I =
1/2$ rule~\cite{HFLAV:2022pw,Manzari:2020eum}, is a long-standing
puzzle.    Various theoretical
models have been proposed to explain this large ratio, but the dual
QCD approach~\cite{Buras:2014maa} and lattice QCD
calculation~\cite{RBC:2020kdj} can only partially explain it.  A
comprehensive understanding of this rule is desirable.

The $\Delta I = 1/2$ rule is also applicable in the decays of spin 1/2
hyperons~\cite{Overseth:1969bxc,Olsen:1970vb}, which can be described 
in terms of its decay asymmetry parameters, $\alpha_{\rm Y}$ and 
$\phi_{\rm Y}$~\cite{Lee:1957qs}. The ratio of decay asymmetry parameters
for the two isospin decay modes $\Lambda\to n\pi^{0}$ and $\Lambda\to p\pi^{-}$,
$\alphaz/\alpham$, is a sensitive probe to determine the contribution
of $\Delta I=3/2$ transitions.  In their absence, the ratio
$\alphaz/\alpham$ is predicted to be unity~\cite{Olsen:1970vb}. A
recent BESIII result suggests that this might not be the
case~\cite{BESIII:2018cnd}.  Further studies of the isospin
amplitude in hyperon decays is required to rigorously test the $\Delta
I = 1/2$ rule.

Moreover, contrary to kaon decays, CP-violation in hyperon decays could arise from
the interference between parity-conserving ($P$-wave) and
parity-violating ($S$-wave) amplitudes with a CP-odd weak phase.
The decay asymmetry parameters of hyperon are CP-odd and assuming CP conservation
$\alpha_{\rm Y} = -\bar{\alpha}_{\rm Y}$ and $ \phi_{\rm Y} =
-\bar{\phi}_{\rm Y}$, where $\bar{\alpha}_{\rm Y}$ and $\bar{\phi}_{\rm Y}$
are decay asymmetry parameters for antihyperon $\bar{Y}$~\cite{Donoghue:1986hh}.
CP symmetry, which is broken in the presence of non-negligible 
weak phase contributions, is gauged by the CP-observables 
$A_{\rm CP}$ and $\Delta\Phi_{\rm CP}$~\cite{BESIII:2021ypr}:

\begin{equation}
A_{\rm CP}^{\rm Y} = \frac{\alpha_{\rm Y} + \bar{\alpha}_{\rm Y}}{\alpha_{\rm Y} - \bar{\alpha}_{\rm Y}}= -\tan \left(\delta_{P}-\delta_{S}\right) \tan \left(\xi_{P}-\xi_{S}\right),
\label{eq:strong}
\end{equation}
\begin{equation}
\Delta \phi_{\rm CP}^{\rm Y} =\frac{\phi_{\rm Y} + \bar{\phi}_{\rm Y}}{2} =\frac{\langle \alpha \rangle}
{\sqrt{1-\langle \alpha \rangle^{2}}} \cos \langle \phi \rangle \tan \left(\xi_{P}-\xi_{S}\right),
\label{eq:weak}
\end{equation}
where $\langle \alpha \rangle = (\alpha_{{\rm Y}} - \bar{\alpha}_{{\rm
    Y}}) /2$, $\langle \phi \rangle = (\phi_{{\rm Y}} -
\bar{\phi}_{{\rm Y}}) /2$, $\delta_{P}-\delta_{S}$ denotes the strong
phase difference of the final-state interaction, and $\xi_P - \xi_S$
denotes the weak phase difference.  Experimentally, the weak phase
difference has been directly determined to be
\mbox{$(1.2\pm3.4\pm0.8)\times10^{-2}$}~rad~\cite{BESIII:2021ypr} for
the decay $\Xi^- \to \Lambda \pi^-$ using entangled $\Xi^-$ and
$\bar{\Xi}^+$ produced at BESIII.

In this Letter, the process $ J/\psi \to \Xi^- \bar{\Xi}^+ \to
\Lambda(p\pi^-) \pi^- \bar{\Lambda}(\bar{n} \pi^0) \pi^+$ is studied
with $(10087\pm44)\times 10^{6}$ $J/\psi$ events~\cite{BESIII:2021cxx}
collected by the BESIII detector.  Benefiting from the
transversely-polarized hyperons and the spin correlation between
hyperon and anti-hyperons, various decay properties of $\Xi^-$ and
$\Lambda$ are determined by an extended formalism that completely
describes the angular distributions of the production and decay
processes~\cite{Perotti:2018wxm}.

The design and performance of the BESIII detector
are described in Refs.~\cite{ablikim2010design, Huang:2022wuo}.  The
corresponding simulation, analysis framework, and software are
presented in Refs.~\cite{Ping:2008zz,Deng:BES,LiW:2006}.  Simulated
Monte Carlo (MC) samples are produced with {\sc
  Geant}4-based~\cite{GEANT4:2002zbu} MC software, which models the
experimental conditions including the electron-positron collision, the
decays of the particles, and the response of the detector. A sample of
simulated events of generic $J/\psi$ decays, corresponding to the
luminosity of data, is used to study the potential background
reactions.  To eliminate experimenter bias, the central values were
blinded by using the hidden answer technique~\cite{Klein:2005di} until
all selections, fits, and uncertainty evaluations were finalized.
Simulated signal and background samples are used to verify the
analysis approaches and to study the systematic effects.  Unless
otherwise indicated, the charge-conjugate channel is implied
throughout the text.

Four charged tracks are required in the multilayer drift
chamber (MDC) within the range $\lvert\cos\theta\rvert< 0.93$, where
$\theta$ is the polar angle with respect to the $z$-axis, which is the
symmetry axis of the MDC.
Due to the non-overlapping momentum ranges of the proton and
pions, a positively charged track with momentum
greater than $0.32$~GeV/$c$ is assigned to be a proton, while a
positively and two negatively charged tracks with momentum less than
$0.30$~GeV/$c$ are assigned to be pions. The probability of
misidentifying a proton for a $\pi^+$ is negligible.  The sequential
decay $\Xi^-\to\Lambda\pi^-\to p\pi^-\pi^-$ is reconstructed by a
vertex fit~\cite{Xu:2010zz,BESIII:2021ypr}, which takes into account the flight paths of the hyperons.
The combination with the smallest $(M_{p\pi^-\pi^-} - m_{\Xi^-})^2 +
(M_{p\pi^-} - m_{\Lambda})^2$ is retained, where
$M_{p\pi^-\pi^-(p\pi^-)}$ denotes the invariant mass of $p\pi^-\pi^-$
($p\pi^-$) and $m_{\Xi^- (\Lambda)}$ refers to the nominal mass of
$\Xi^-$ ($\Lambda$)~\cite{Workman:2022ynf}.  The probability of a
$\pi^-$ from the $\Lambda$ and $\Xi^-$ decays being wrongly assigned
is found to be $0.1\%$, which is negligible.  The candidate events are
required to satisfy $\lvert M_{p\pi^-} - m_{\Lambda}\rvert <
11$~MeV/$c^2$ and $\lvert M_{p\pi^-\pi^-} - m_{\Xi^-}\rvert <
11$~MeV/$c^2$.  The decay lengths of the $\Xi^-$ and $\Lambda$ are
calculated in the vertex fit and required to be positive.  To improve
the resolution and minimize the discrepancy between data and MC
simulation, the polar angle $\theta_{\Xi^-}$ of the reconstructed $\Xi^-$
in the $e^+e^-$ center-of-mass frame is required to satisfy
$\lvert\cos\theta_{\Xi^-}\rvert < 0.84$.

At least two photon candidates in the electromagnetic calorimeter
(EMC) are required.  A photon candidate should have energy greater
than $25$~MeV in the barrel region ($\lvert\cos\theta\rvert < 0.8$) or
$50$~MeV in the end-cap region ($0.86 < \lvert\cos\theta \rvert <
0.92$).  For antiprotons, which may annihilate in the detector, photon
candidates must be separated from charged tracks with an opening angle
greater than $20^\circ$, while for other tracks the angle must be
greater than $10^\circ$.  To suppress electronic noise and showers
unrelated to the event, photon candidates are required to have the
EMC time difference from the event start time within $[0, 700]$~ns.
To veto the showers from antineutron interactions in the EMC, the
photon candidates should be separated from the direction of the $\Xi^-
\pi^+$ recoiling system with an opening angle greater than $15^\circ$.
A boosted decision tree (BDT) classifier~\cite{Therhaag:2010zz} is constructed based on the
shower shape variables to discriminate a signal photon from a noise
shower.  The shower shape variables include the deposited energy,
number of hits, second and Zernike moments, and deposition
shape~\cite{cluster:shape}.  The signal efficiency of the BDT is
$90\%$, and $55\%$ of the background is rejected.  The $\pi^0$
candidates are reconstructed from a pair of photons by constraining
their invariant mass to the $\pi^0$ nominal mass, and the
corresponding $\chi^2_{1C}$ is required to be less than 25.  Due to
combinatorial effects, it is possible to have more than one unique
$\pi^0$ candidate in a single event.

\begin{figure}[hpt]
  \centering
  \mbox
  {
  	\begin{overpic}[width=0.45\textwidth]{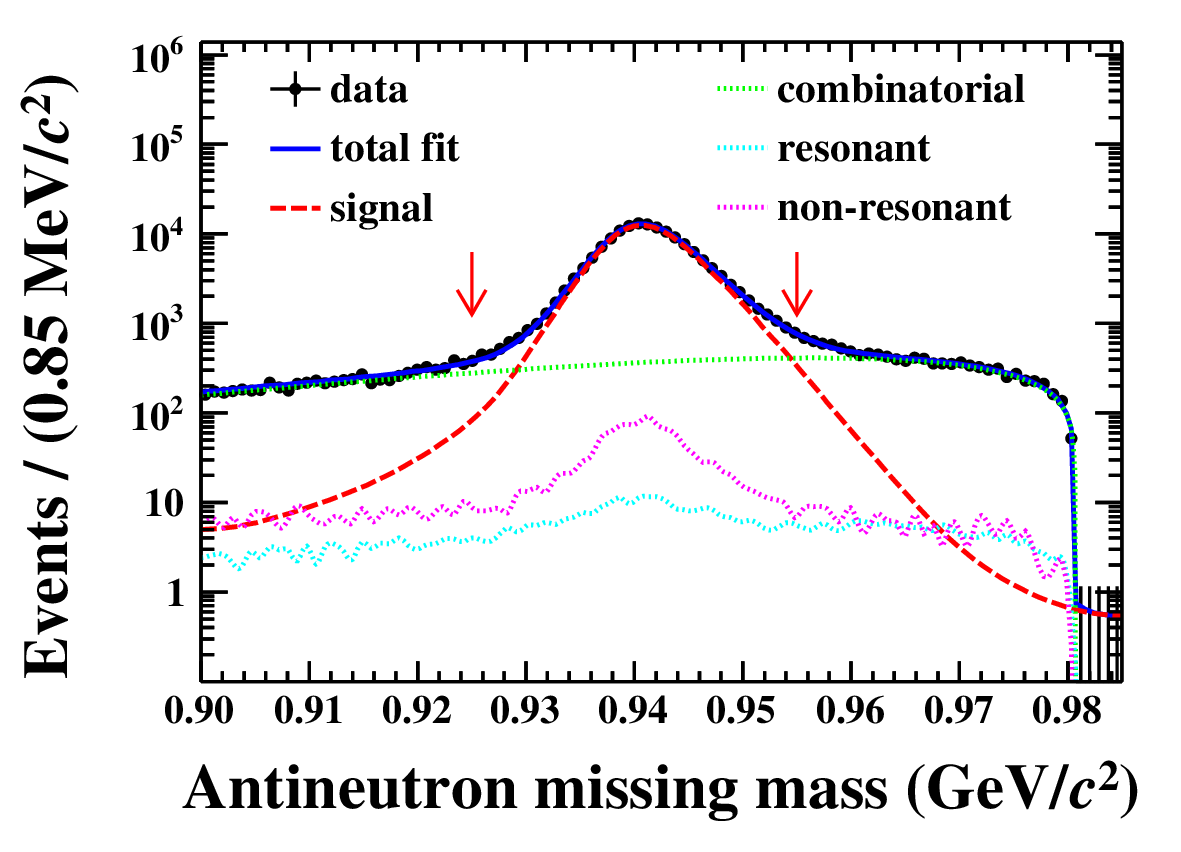}
  	\end{overpic}
  }
\caption{Distribution of antineutron missing mass. 
The data are shown as black data points with error bars.
The blue solid curve represents the total fit result,
and the red dashed line denotes the signal shape.
The dotted lines in green, light blue and magenta denote the combinatorial, 
resonant and non-resonant background contributions, respectively. 
The red arrows indicate the signal region.}
 \label{fig:hyperondecay}
\end{figure}

A kinematic fit under the hypothesis of $J/\psi \to \Xi^-
\pi^+\bar{n}\gamma\gamma$ is performed imposing energy-momentum
conservation and constraining the invariant masses of $\gamma\gamma$
and $\gamma\gamma\bar{n}$ to the nominal masses of $\pi^0$ and
$\bar{\Lambda}$, respectively.  The kinematics of the $\Xi^-$ are obtained
from the above vertex fit.  The antineutron is treated as a missing
particle with unknown mass.  The fit is performed for each $\pi^0$
candidate.  If there is more than one $\pi^0$ candidate, the candidate
with the smallest $\chi^2$ is retained, and $\chi^2 < 200$ is
required.  The invariant mass of $\bar{n}\gamma\gamma\pi^+$ is
required to satisfy $\lvert M_{\bar{n}\gamma\gamma\pi^+} -
m_{\bar{\Xi}^+}\rvert < 11$~MeV/$c^2$.  Since all other final
state particles are detected, the kinematic fit allows for the reconstruction of
the four momentum of antineutron.  The signal is identified by
the antineutron's missing mass, as shown in Fig.~\ref{fig:hyperondecay} with a
prominent signal peak in the antineutron vicinity and a low level
background.

Detailed studies are performed with MC simulation and data in the
$\Xi^-$ and $\bar{\Xi}^+$ sideband regions to evaluate the potential
backgrounds.  The dominant background,  referred to as
combinatorial background, is from signal events with
misreconstructed $\pi^0$ candidates, which does not peak in the
antineutron missing mass distribution.  The remaining background sources
are classified into two categories~\cite{Zhou:2020ksj}: resonant
background that contains $\Xi^-\bar{\Xi}^+$ intermediate states, such as,
$J/\psi \to \gamma\eta_c \to \gamma \Xi^-\bar{\Xi}^+ \to \gamma
\Lambda(\to p\pi^-)\pi^- \bar{\Lambda}(\bar{n}\pi^0)\pi^+$ and
$J/\psi\to \Xi^-\bar{\Xi}^+ \to \Lambda(\to p\pi^-)\pi^-
\bar{\Lambda}(\bar{p}\pi^+)\pi^+$; non-resonant background without
$\Xi^{-}\bar{\Xi}^{+}$ intermediate states.  The decay processes of
resonant backgrounds are well understood, and the corresponding
contributions are evaluated by MC simulation, which are generated
according to the helicity amplitudes and weighted according to the
branching fractions~\cite{Workman:2022ynf}.  MC simulation shows that
the distributions of $M_{p\pi^-\pi^-}$ and
$M_{\bar{n}\gamma\gamma\pi^+}$ of non-resonant background are almost
flat. Therefore, the corresponding contribution can be evaluated from
the $\Xi^-$ and $\bar{\Xi}^+$ sideband regions.

Signal yields are obtained from an unbinned maximum likelihood fit of
the missing mass distribution.  In the fit shown in
Fig.~\ref{fig:hyperondecay}, the signal is described by an
MC-simulated shape convolved by a Gaussian function accounting for the
resolution difference between data and MC simulation.  The
combinatorial background is described by the signal MC sample, and it
is parameterized by a product of an ARGUS
function~\cite{ARGUS:1990hfq} and a cubic function.  Fixing both
the magnitude and shape, the resonant and non-resonant backgrounds are
described with the MC simulation and data events in the sideband
region, respectively.  The normalization of the background and the
definition of the sideband region are shown in Sec.~2 of the
Supplemental Material~\cite{supp:systematic}.  The fit yields $143973
\pm 414$ signal events and a purity of $91.2\%$ in the mass range $[0.925,
  0.955]$~GeV/$c^2$.  The same procedure is performed for the
charge-conjugate process and results in $123208 \pm 382$ signal
events and a purity of $91.0\%$.


The joint angular amplitude of the full decay chain can be written in
a modular form as
\begin{equation}
\mathcal{W}(\xi ; \omega)=\sum_{\mu, \nu=0}^{3} C_{\mu \nu} \sum_{\mu^{\prime}, \nu^{\prime}=0}^{3} a_{\mu \mu^{\prime}}^{\Xi} a_{\nu \nu^{\prime}}^{\Xi} a_{\mu^{\prime} 0}^{\Lambda} a_{\nu^{\prime} 0}^{\bar{\Lambda}}.
\end{equation}
Here $C_{\mu \nu}$ is a $4\times4$ real-valued spin density matrix
describing the spin configuration of the entangled $\Xi^- \bar{\Xi}^+$
pair, $a^{{\rm Y}}_{\mu \nu}$ is also a $4\times4$ real-valued matrix
representing the propagation of the spin density matrix in the decays
of a spin 1/2 hyperon into a spin 1/2 baryon and a pseudoscalar, ${\rm
  Y}\to {\rm B}\pi$.  Therefore, the distribution of the nine helicity
angles $\xi = (\theta_{\Xi}, \theta_{\Lambda}, \phi_{\Lambda},
\theta_{\bar{\Lambda}}, \phi_{\bar{\Lambda}}, \theta_{p}, \phi_{p},
\theta_{\bar{n}}, \phi_{\bar{n}})$ is determined by eight global
parameters \mbox{$\omega = (\alpha_{J/\psi}, \Delta\Phi_{J/\psi},
  \alphaXi, \phi_{\Xi}, \bar{\alpha}_{\Xi}, \bar{\phi}_{\Xi}, \alpham,
  \baralphaz)$}.  In this analysis, ${\rm Y}\to {\rm B}\pi$ stands for
$\Xi^-\to \Lambda \pi^-$, $\Lambda\to p\pi^-$ and $\bar{\Lambda}\to \bar{n}\pi^0$.
The distribution of the helicity angle $\theta_{p}$ in the $\Lambda$
rest frame is written as
\begin{equation}
\label{eq:phel}
	\frac{\mathrm{d} N}{\mathrm{d} \cos\theta_p} \propto 1 + \alpham \alphaXi \cos\theta_{p}
\end{equation}
by integrating over the remaining eight helicity angles.
The formalism of the full angular distribution and the definition of the reference system are discussed in detail in Ref.~\cite{BESIII:2021ypr}.

A simultaneous fit on the joint angular distribution is carried out
with the production parameters, $\alpha_{J/\psi}$ and $\Delta\Phi_{J/\psi}$, and decay asymmetry parameters of
$\Xi^-$ shared between the two charge-conjugate channels.  For each
channel, the probability distribution function of the eight unknown
parameters $\omega$ can be defined in terms of the helicity angles
$\xi$
\begin{equation}
	\mathcal{P}(\xi;\omega) = \mathcal{W}(\xi;\omega)\varepsilon(\xi)/\mathcal{N}(\omega),
\end{equation}
where the normalization factor $\mathcal{N}(\omega)$ is calculated with  $\mathcal{N}(\omega) = \frac{1}{M} \sum_{j = 1}^{M} \frac{\mathcal{W}(\xi_j;\omega)}{\mathcal{W}(\xi_j;\omega_{\rm gen})}$ by a signal MC sample generated with parameters $\omega_{\rm gen}$.
The sum runs over all events in the generated sample $M$, which is chosen to be thirty times the yield 
obtained in data after the full selection.
The log-likelihood function for $N$ observed events is
\begin{equation}
	\small
	\mathcal{S} = - \mathcal{G} \left( \sum_{i = 1}^N\ln \mathcal{P}(\xi_i;\omega) -
	\sum_j {\sc w_j} \sum_i^{N_j^{\rm bkg}}\ln \mathcal{P} (\xi_i ; \omega ) \right),
\end{equation}
where the second term in brackets with $j$ from one to three
represents the three different sources of background remaining in the
final event sample.  Their contributions are evaluated with the
corresponding MC samples or data events in the sideband region and
their associated weight factors $w_j$.  The global factor,
\mbox{$\mathcal{G} = (N - \sum_j N_j^{\rm bkg} \times \sc{w_j})/ (N +
  \sum_j N_j^{\rm bkg}\times \sc{w_j}^2)$} , corrects for the statistical
uncertainties in the weighted likelihood
fit~\cite{Langenbruch:2019nwe}.

The $\mathcal{S}$ function is minimized using
Minuit2~\cite{Hatlo:2005cj} to determine the production and
decay asymmetry parameters $\omega$.  The results from the
fit, as shown in Table~\ref{tab:fitresults}, are consistent with
previous measurements, but with improved precision.  In particular,
$\alphaz$ is almost the same in magnitude and opposite in sign as
$\baralphaz$, and its precision is improved by a factor of four over
previous measurements.

\begin{table}[hbtp]
	\centering
	\scriptsize
	\caption[]{The production and decay asymmetry parameters, the
          weak and strong phase differences from $\Xi^-$ decay, the
          tests of CP symmetry, and the ratios of decay asymmetry
          parameters, $\alphaz/\alpham$ and $\baralphaz/\alphap$. The
          first and second uncertainties are statistical and
          systematic, respectively.}  \renewcommand\arraystretch{1.5}
	\label{tab:fitresults}
 \resizebox{\linewidth}{!}{
	\begin{tabular}{lrr}
		\hline \hline
		Parameters & \makecell[c]{This work} & \makecell[c]{Previous result}   \\
		\hline
$\alpha_{J/\psi}$ & $ 0.611 \pm 0.007 ^{+0.013}_{-0.007}$& $0.586\pm0.012\pm0.010$~\cite{BESIII:2021ypr}   \\
$\Delta\Phi_{J/\psi}$ (rad) & $ 1.30 \pm 0.03 ^{+0.02}_{-0.03}$ & $1.213\pm0.046\pm0.016$~\cite{BESIII:2021ypr}   \\
$\alpha_{\Xi}$ & $ -0.367 \pm 0.004 ^{+0.003}_{-0.004}$& $-0.376\pm0.007\pm0.003$~\cite{BESIII:2021ypr}   \\
$\phi_{\Xi}$ (rad) & $ -0.016 \pm 0.012 ^{+0.004}_{-0.008}$ &  $0.011\pm0.019\pm0.009$~\cite{BESIII:2021ypr}  \\
$\bar{\alpha}_{\Xi}$ & $ 0.374 \pm 0.004 ^{+0.003}_{-0.004}$& $0.371\pm0.007\pm0.002$~\cite{BESIII:2021ypr}  \\
$\bar{\phi}_{\Xi}$ (rad) & $ 0.010 \pm 0.012 ^{+0.003}_{-0.013}$ & $-0.021\pm0.019\pm0.007$~\cite{BESIII:2021ypr}  \\
$\alpham$ & $ 0.764 \pm 0.008 ^{+0.005}_{-0.006}$& $0.7519\pm0.0036\pm0.0024$~\cite{BESIII:2022qax}   \\
$\alphap$ & $ -0.774 \pm 0.009 ^{+0.005}_{-0.005}$& $-0.7559\pm0.0036\pm0.0030$~\cite{BESIII:2022qax}   \\
$\alphaz$ & $ 0.670 \pm 0.009 ^{+0.009}_{-0.008}$& $0.75\pm0.05$~\cite{Workman:2022ynf}   \\
$\baralphaz$ & $ -0.668 \pm 0.008 ^{+0.006}_{-0.008}$& $-0.692\pm0.016\pm0.006$~\cite{BESIII:2018cnd}   \\
\hline
$\delta_{P} - \delta_{S}$ (rad) & $ 0.033 \pm 0.020 ^{+0.008}_{-0.012}$ & $-0.040\pm0.033\pm0.017$~\cite{BESIII:2021ypr}   \\
$\xi_{P} - \xi_{S}$ (rad) & $ 0.007 \pm 0.020 ^{+0.018}_{-0.005}$ & $0.012\pm0.034\pm0.008$~\cite{BESIII:2021ypr}  \\
\hline
$A_{\rm CP}^{\Xi}$ & $ -0.009 \pm 0.008 ^{+0.007}_{-0.002}$ &  $0.006\pm0.013\pm0.006$~\cite{BESIII:2021ypr}    \\
$ \Delta \phi_{\rm CP}^{\Xi}$ (rad) & $ -0.003 \pm 0.008 ^{+0.003}_{-0.007}$ & $-0.005\pm0.014\pm0.003$~\cite{BESIII:2021ypr}    \\
$A_{\rm CP}^{-}$ & $ -0.007 \pm 0.008 ^{+0.002}_{-0.003}$ & $-0.0025\pm0.0046\pm0.0012$~\cite{BESIII:2022qax}    \\
$A_{\rm CP}^{0}$ & $ 0.001 \pm 0.009 ^{+0.005}_{-0.007}$ & -    \\
$A_{\rm CP}^{\Lambda}$ & $ -0.004 \pm 0.007 ^{+0.003}_{-0.004}$ & -    \\
		\hline
		$\alphaz/\alpham$ & $ 0.877 \pm 0.015 ^{+0.014}_{-0.010}$ & $1.01 \pm 0.07$~\cite{Workman:2022ynf}    \\
$\baralphaz/\alphap$ & $ 0.863 \pm 0.014 ^{+0.012}_{-0.008}$ & $0.913 \pm 0.028 \pm 0.012$~\cite{BESIII:2018cnd}    \\
		\hline \hline
	\end{tabular}
 }
\end{table}

If CP is conserved, the product of the decay asymmetry parameters
$\alpham \alpha_{\Xi}$ and $\alphap \bar{\alpha}_{\Xi}$ should be
equal to each other, and the ratios of helicity angular distributions
for different nucleons in the final states, $R(\cos \theta_{p}, \cos
\theta_{\bar{p}}) = (1 + \alpham \alpha_{\Xi} \cos\theta_{p}) / (1 +
\alphap \bar{\alpha}_{\Xi} \cos\theta_{\bar{p}})$ and $R(\cos
\theta_{n}, \cos \theta_{\bar{n}}) = (1 + \alphaz \alphaXi
\cos\theta_{n}) /( 1 + \baralphaz \bar{\alpha}_{\Xi}
\cos\theta_{\bar{n}})$, are flat and equal to unity. In a similar way,
if there is no $\Delta I = 3/2$ transition in $\Lambda$ decay,
$\alpham$ should be equal to $\alphaz$ and the ratios, $R(\cos
\theta_{n}, \cos \theta_{p}) = (1 + \alphaz \alpha_{\Xi}
\cos\theta_{n}) /( 1 + \alpham \alpha_{\Xi} \cos\theta_{p})$ and
$R(\cos \theta_{\bar{n}}, \cos \theta_{\bar{p}}) = (1 + \baralphaz
\bar{\alpha}_{\Xi} \cos\theta_{\bar{n}}) / (1 + \alphap
\bar{\alpha}_{\Xi} \cos\theta_{\bar{p}})$, are also flat and equal to
unity. The accuracy of the CP symmetry and the $\Delta I = 1/2$ rule
tests can be improved by using the isospin average for $R_1$, $R_1 =
(1+ \alpha_{\Lambda} \alphaXi \cos \theta)/( 1+ \bar{\alpha}_{\Lambda}
\bar{\alpha}_{\Xi} \cos \theta )$, where $\cos\theta$ stands for 
the helicity angle of nucleon, $\alpha_{\Lambda}$ is defined
as $ (2 \alpham + \alphaz) / 3$, and the average of the decay symmetry
parameters of hyperon and antihyperon for $R_2$, $R_2 = (1+\langle \alphaz
\rangle \langle \alphaXi \rangle \cos \theta)/( 1+\langle \alpham
\rangle \langle \alphaXi \rangle \cos \theta )$.

\begin{figure}[hpt]
  \centering
  \mbox
  {
  \begin{overpic}[width=0.45\textwidth]{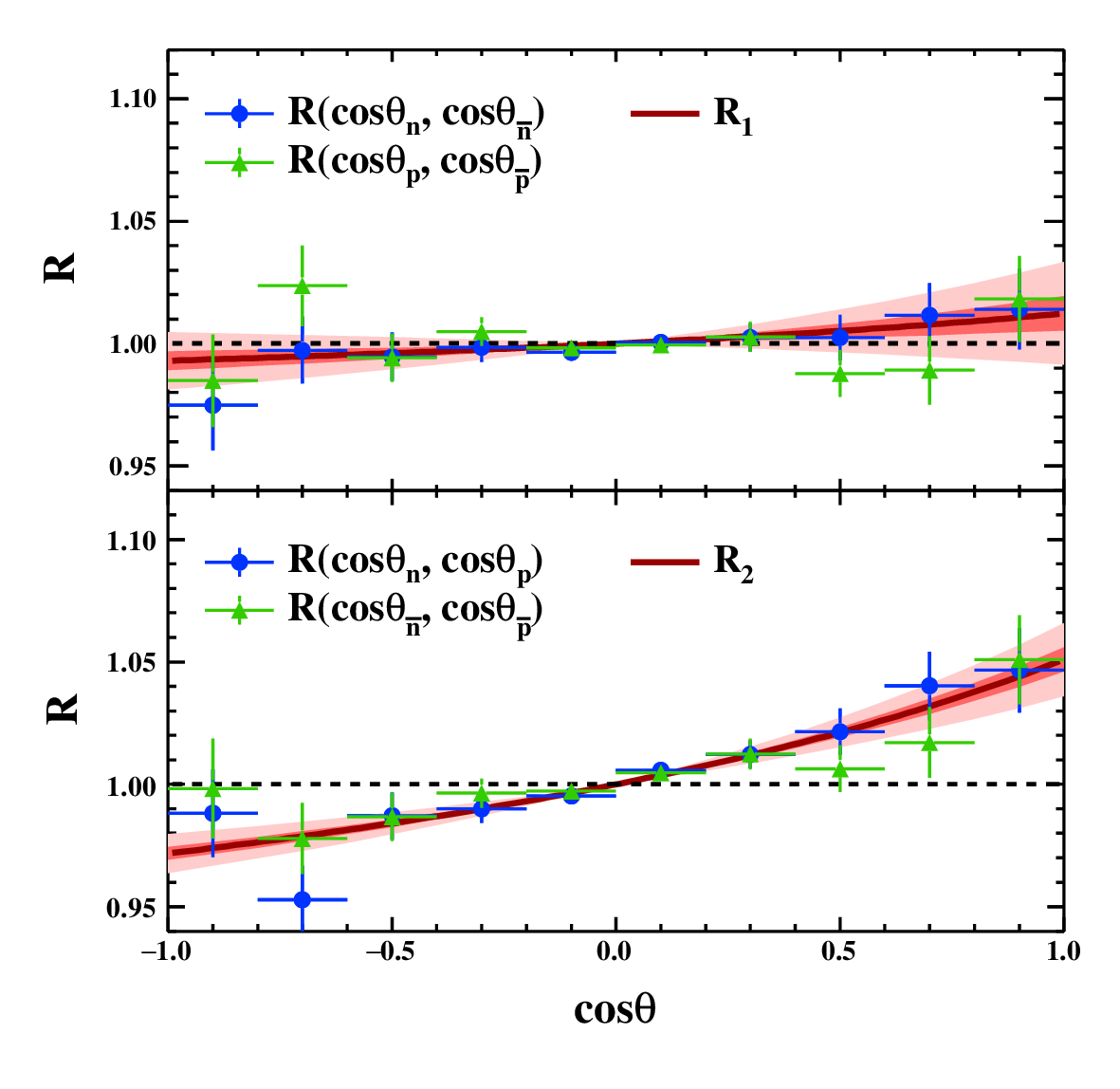}
  \end{overpic}
  } 

\caption{ The ratios of helicity angular
  distributions for different nucleons in the final states,
	$R(\cos \theta_{p}, \cos \theta_{\bar{p}})$ and $R(\cos \theta_{n}, \cos \theta_{\bar{n}})$ (top) as well as
	$R(\cos \theta_{n}, \cos \theta_{p})$ and $R(\cos \theta_{\bar{n}}, \cos \theta_{\bar{p}})$ (bottom) versus $\cos \theta$.
The dots with errors are determined by independent fits for each $\cos \theta$ bin of the corresponding nucleons.
	The solid curves in red with 1$\sigma$ (red) and 3$\sigma$ (pink) statistical uncertainty bands show the results of the simultaneous fit.
	The dashed curves in black show the CP-conserving and no $\Delta I = 3/2$ transition expectations.
 }
 \label{fig:ratio}
\end{figure}

To illustrate the tests of CP symmetry and the $\Delta I = 1/2$ rule,
four ratios of the helicity angular distributions for different
nucleons in the final states are shown in Fig.~\ref{fig:ratio} by
dots with error bars. $R_1$ and $R_2$ with parameters
from Table~\ref{tab:fitresults} are also presented in
Fig.~\ref{fig:ratio}.  The ratios obtained by fitting the events in
bins of $\cos \theta$ are in good agreement with the global curves
obtained for $R_1$ and $R_2$.  The nearly flat distribution of $R_1$
is consistent with CP conservation. The sloping distribution
of $R_2$ indicates the existence of the
contribution of $\Delta I = 3/2$ transition in $\Lambda$ decay.

The systematic uncertainties are split into different categories:
reconstruction and event selection of the signal candidates, the
uncertainties related to the background contributions, and the effects
which arise from the final fit procedure.  The uncertainty of the
$\pi^0$ reconstruction is investigated by studying the decay $J/\psi
\to \Sigma^+ (p\pi^0) \pi^- \bar{\Lambda} (\bar{p} \pi^+) + c.c.$ as
it has a similar final state topology and decay length as the signal.
The systematic uncertainty from $\pi^\pm$ reconstruction is
investigated by using a control sample of $J/\psi \to \Xi^-\bar{\Xi}^+
\to \Lambda (p\pi^-) \pi^- \bar{\Lambda} (\bar{p} \pi^+)\pi^+ + c.c.$.
The systematic uncertainties related to the selection criteria (the
decay points and invariant masses of $\Lambda$ and $\Xi^-$, the polar
angle of $\Xi^-$, the missing mass and the $\chi^2$ of the kinematic
fit) are studied by varying their required values around the nominal
ones and repeating the fit.  The uncertainty due to the combinatorial
background is determined by both smearing the parameters of model and
varying its yield from the fit to the missing mass distribution by $\pm 1\sigma$. 
The uncertainties associated with the resonant backgrounds, which are
propagated from the uncertainties in branching fractions, number of
$J/\psi$ events and MC sample statistics, are also evaluated by
varying the background yield by $\pm 1\sigma$. In the case of
non-resonant background, the fit is repeated without this background
component, and the deviation from the nominal fit is taken as the
systematic uncertainty.  To estimate the systematic uncertainty of the
fit procedure, 1000 sets of toy MC samples are generated with the
parameters from Table~\ref{tab:fitresults}.  Each set is fitted to
obtain the distribution of the output parameters.  
The average values of the 
difference between the input and output parameters
and statistical errors of the average differences
are regarded as systematic uncertainties. More details can be found in the
Supplemental Material~\cite{supp:systematic}.

In summary, the decay asymmetry parameters listed in
Table~\ref{tab:fitresults} are simultaneously determined from the
process $J/\psi \to \Xi^- \bar{\Xi}^+ \to \Lambda(p\pi^-) \pi^-
\bar{\Lambda}(\bar{n} \pi^0) \pi^+$ and its charge-conjugate channel
with $(10087\pm44)\times 10^{6}$ $J/\psi$ events collected by the
BESIII detector.  Using Eq.~\ref{eq:strong} and Eq.~\ref{eq:weak}, the
CP observables $A_{\rm CP}^{\Xi}$ and $\Delta \phi_{\rm CP}^{\Xi}$ for $\Xi^-$
decay, as well as $A_{\rm CP}^- = (\alpham + \alphap)/(\alpham - \alphap)$ 
and $A_{\rm CP}^0 = (\alphaz + \baralphaz)/(\alphaz - \baralphaz)$ of the charged and neutral
$\Lambda$ decays, are obtained from the corresponding decay asymmetry
parameters and correlations.  $A_{\rm CP}^{\Xi}$ and $\Delta
\phi_{\rm CP}^{\Xi}$ are measured with world-leading precision, and
$A_{\CP}^0$ is measured for the first time.  The correlations $\rho
(\alpham,\alphap)$ and $ \rho (\alphaz, \baralphaz)$ measured from two
charge-conjugate channels are negligible.  The precise CP
symmetry test of the $\Lambda$ decay is conducted with its isospin
averages, $A_{\rm CP}^{\Lambda} = (2 A_{\CP}^- + A_{\CP}^0)/3$, which
improves the sensitivity of the CP symmetry test by 20\% compared to the
individual tests for each isospin decay mode.  The strong phase and
weak phase differences of $\Xi^- \to \Lambda \pi^-$, derived from
Eq.~\ref{eq:strong} and Eq.~\ref{eq:weak}, are both consistent with
previous BESIII results~\cite{BESIII:2021ypr}.  The strong phase
difference is also in agreement with the HyperCP
measurement~\cite{HyperCP:2004zvh}.  The CP symmetry is conserved in
the decay of $\Xi^-$ and $\Lambda$ within the current precision.  The
theoretical predictions within the Standard
Model~\cite{tandean2003cp,He:2022bbs} are $0.5 \times 10^{-5}
\leqslant (A_{\rm CP}^{\Xi} )_{\mathrm{SM}} \leqslant 6 \times 10^{-5}$,
$-3.8 \times 10^{-4} \leqslant(\xi_{P}-\xi_{S})_{\mathrm{SM}}
\leqslant-0.3 \times 10^{-4}$ and $-3 \times 10^{-5} \leqslant(A_{C
  P}^{\Lambda})_{\mathrm{SM}} \leqslant 3 \times 10^{-5}$.

The ratios of $\alphaz/\alpham$ and $\baralphaz/\alphap$ deviate from
unity by more than $5$ standard deviations, which signifies the
existence of the $\Delta I = 3/2$ transition in both $\Lambda$ and
$\bar{\Lambda}$ decays for the first time.  
Using the averages of the ratio $\langle
\alphaz \rangle / \langle \alpham \rangle = 0.870 \pm 0.012
^{+0.011}_{-0.010}$ with combinations of the decay rates $\Gamma(\Lambda \to
p\pi^-)$, $\Gamma(\Lambda \to n\pi^0)$~\cite{Workman:2022ynf} and the
$N$-$\pi$ scattering phase shift~\cite{Hoferichter:2015hva}, the ratio
of $\Delta I = 3/2$ to $\Delta I = 1/2$ transitions in $S$-wave is
determined to be $S_1/S_3 = 28.4 \pm 1.3 ^{+1.1}_{-1.0} \pm 3.9$, while
in $P$-wave $P_1/P_3 = -13.0 \pm 1.4 ^{+1.1}_{-1.2} \pm 0.7$
according to Ref.~\cite{Salone:2022lpt}, where the first uncertainties
are statistical, the second systematic and the third from the input
parameters.  The ratio in $S$-wave is consistent  
with ${\rm Re}(A_0)/{\rm Re}(A_2)$ in $K \to \pi\pi$ within the uncertainty, 
while the ratio in $P$-wave is measured for the first time
and found different from that in $S$-wave.
This measurement provides a
constraint for lattice QCD~\cite{RBC:2020kdj} and dual
QCD~\cite{Buras:2014maa} approach to understand the $\Delta I = 1/2$ rule.


The authors thank Professor X.G. He and Professor X. Feng for helpful discussions.
The BESIII Collaboration thanks the staff of BEPCII and the IHEP computing center and the supercomputing  center of the University of Science and Technology of China (USTC)  for their strong support. This work is supported in part by National Key R\&D Program of China under Contracts Nos. 2020YFA0406300, 2020YFA0406400; National Natural Science Foundation of China (NSFC) under Contracts Nos. 11635010, 11735014, 11835012, 11935015, 11935016, 11935018, 11961141012, 12022510, 12025502, 12035009, 12035013, 12061131003, 12192260, 12192261, 12192262, 12192263, 12192264, 12192265, 12221005, 12225509, 12235017, 12122509, 12105276, 11625523; the Chinese Academy of Sciences (CAS) Large-Scale Scientific Facility Program; the CAS Center for Excellence in Particle Physics (CCEPP); Joint Large-Scale Scientific Facility Funds of the NSFC and CAS under Contract No. U1832207, U2032111, U1732263, U1832103; CAS Key Research Program of Frontier Sciences under Contracts Nos. QYZDJ-SSW-SLH003, QYZDJ-SSW-SLH040; 100 Talents Program of CAS; The Institute of Nuclear and Particle Physics (INPAC) and Shanghai Key Laboratory for Particle Physics and Cosmology; European Union's Horizon 2020 research and innovation programme under Marie Sklodowska-Curie grant agreement under Contract No. 894790; German Research Foundation DFG under Contracts Nos. 455635585, Collaborative Research Center CRC 1044, FOR5327, GRK 2149; Istituto Nazionale di Fisica Nucleare, Italy; Ministry of Development of Turkey under Contract No. DPT2006K-120470; National Research Foundation of Korea under Contract No. NRF-2022R1A2C1092335; National Science and Technology fund of Mongolia; National Science Research and Innovation Fund (NSRF) via the Program Management Unit for Human Resources \& Institutional Development, Research and Innovation of Thailand under Contract No. B16F640076; Polish National Science Centre under Contract No. 2019/35/O/ST2/02907; The Swedish Research Council; U. S. Department of Energy under Contract No. DE-FG02-05ER41374; Olle Engkvist Foundation under Contract No. 200-0605 and Lundström-Åman Foundation.


\bibliographystyle{prlapsrev4-1}
\bibliography{apssampbib}

\end{document}